\documentclass[12pt]{article}
\usepackage[utf8]{inputenc}
\usepackage{setspace}
\usepackage[linesnumbered,ruled]{algorithm2e}
\usepackage{latexsym}
\usepackage{graphicx}
\usepackage{amsmath}
\usepackage{multirow}
\usepackage{natbib}
\usepackage{color}
\usepackage{comment} 
\usepackage{bbding}
\usepackage{pifont}
\usepackage{amssymb}
\usepackage{amsthm,amsmath,amsfonts}

\usepackage{array}
\newcolumntype{H}{>{\setbox0=\hbox\bgroup}c<{\egroup}@{}}

\newcommand{ \bzero}{\mathbf{0}}

\newcommand{\bphi}{{\boldsymbol{\phi}}}
\newcommand{\bbeta}{{\boldsymbol{\beta}}}
\newcommand{\bgamma}{{\boldsymbol{\gamma}}}

\newcommand{\bbg}{{\boldsymbol{g}}}

\newcommand{\bI}{{\boldsymbol{I}}}

\newcommand{\bV}{{\boldsymbol{V}}}
\newcommand{\blambda}{\mbox{\boldmath$\lambda$}}

\newcommand{\en}{\end{equation}}
\newcommand{\bea}{\begin{eqnarray}}
\newcommand{\eea}{\end{eqnarray}}

\newcommand{\ea}{\end{array}}

\newcommand{\bx}{{\boldsymbol{x}}}
\newcommand{\bX}{{\boldsymbol{X}}}

\newcommand{\bpi}{{\boldsymbol{\pi}}}

\newcommand{\bean}{\begin{eqnarray*}}
\newcommand{\eean}{\end{eqnarray*}}

\def\T{{ \mathrm{\scriptscriptstyle T} }}

\title{Multiple bias-calibration for adjusting selection bias of non-probability samples using data integration}
 \author{ Z. Wang \and S. Yang \and J.K. Kim}
\date{}

\begin{document}

\maketitle

\begin{abstract}
Valid statistical inference is challenging when the sample is subject to unknown selection bias. Data integration can be used to correct for selection bias when we have a parallel probability sample from the same population with some common measurements. How to model and estimate the selection probability or the propensity score (PS) of a non-probability sample using an independent probability sample is the challenging part of the data integration. We approach this difficult problem by employing multiple candidate models for PS combined with empirical likelihood. By incorporating multiple propensity score models into the internal bias calibration constraint in the empirical likelihood setup, the selection bias can be eliminated so long as the multiple candidate models contain a true PS model. The bias calibration constraint under the multiple PS models is called multiple bias calibration. Multiple PS models can include both missing-at-random and missing-not-at-random models. Asymptotic properties are discussed, and some limited simulation studies are presented to compare the proposed method with some existing competitors. Plasmode simulation studies using the Culture \& Community in a Time of Crisis dataset demonstrate the practical usage and advantages of the proposed method. 

\vspace{0.5cm}
\end{abstract}

\doublespacing
\newpage 

\section{Introduction}
While probability samples serve as the gold standard in social science and related fields to estimate population quantities and monitor policy effects, their acquisition and analysis become challenging due to their high cost and low response rates \citep{couper2013sky,miller2017there,williams2018trends,kalton2019developments}. During the past 20 years, non-probability samples, on the other hand, have been increasingly prevailing since they are ``wider, deeper, faster, better, cheaper, less burdensome, and more relevant'' \citep{holt2007official,citro2014multiple}. However, non-probability samples suffer from selection bias because of unknown selection probabilities, and it may lead to erroneous inference if such selection bias is overlooked \citep{couper2000web,bethlehem2016solving,elliott2017inference,meng2018statistical}. Consequently, addressing selection bias in non-probability samples constitutes a significant research area, and we tackle this challenge through the utilization of a data integration methodology.

Data integration is an emerging area of research that combines multiple data sources in a defensible way. 
\cite{tam15} and \cite{pfeffermann15} addressed the methodological challenges of big data in the production of official statistics. However, important questions related to data integration for complex surveys remain, as summarized by \cite{rao2020} and \cite{yang2020}.
In data integration, by utilizing an independent probability sample as a calibration sample, the selection bias in the non-probability  sample can be reduced.   \citet{valliant2011estimating} proposed a rescaled design weight (RDW) model based on a missing-at-random (MAR) assumption, and the associated model parameter is estimated by combining both probability sample and non-probability sample. Based on the estimated selection probabilities for the non-probability sample, \citet{valliant2011estimating} used a H{\'a}jek estimator \citep{hajek1971comment} to estimate the parameter of interest. Similarly, \citet{chen2020doubly} proposed a parametric MAR model for the selection probabilities, but the model parameters are estimated by balancing a smooth function involving them. \citet{wang2021adjusted} proposed several estimators to correct the selection bias of the non-probability sample based on different logistic regression  models. \citet{kimtam2021} proposed an efficient estimator using a regression model to integrate information from a probability sample and a non-probability sample. Parametric models are commonly used by existing methods, but they are sensitive to model misspecification. \citet{rivers2007sampling} proposed a nearest-neighbor sampling matching approach, which is a non-parametric imputation method for data integration. Semi-parametric imputation is also discussed by \citet{kim2021combining}. \citet{wang2022functional} proposed a non-parametric data integration method based on models from a reproducing kernel Hilbert space. \citet{chen2022pseudo} proposed a  data integration method for a non-probability sample by empirical likelihood \citep{owen1991empirical}.  See also \citet{wu2022statistical}, \citet{beaumont2020probability} and the references therein for other data integration methods. Existing methods mainly assumed an MAR condition for the selection probabilities of the non-probability sample, and such an assumption is hard to verify in practice. In addition, violation of the MAR assumption usually leads to biased inference. 



In this paper, we propose a maximum empirical likelihood (MEL) estimator to integrate a probability sample, a non-probability sample and fully observed auxiliary information from the population. Differently from most existing work, we accommodate a possible  missing-not-at-random (MNAR)  model for the selection probability associated with the non-probability sample. Instead of assuming a correctly specified parametric  model, we  assume that a true response model for the selection probability belongs to a set of candidate models for the selection probability inspired by the multiple robustness framework of \citet{han2013}. The price we need to pay for the MNAR assumption is that a correctly specified density function is required. The assumption on the density function is less stringent than the MAR one in the sense that the former can be checked or obtained by the probability sample but the latter one cannot. \citet{han2013} did not provide a rigorous variance estimator, but under regularity conditions,  an asymptotically unbiased variance estimator is available for the proposed estimator. 

The remainder of the paper is organized as follows. The problem setup is introduced in Section~\ref{sec: problem setup}. We propose the maximum empirical likelihood estimator in Section~\ref{sec: MELE}, and its variance estimator is discussed in Section~\ref{sec: var est}. In Section~\ref{sec: simu section}, the proposed estimator is compared with the existing ones in a limited simulation study. The application of the proposed estimator to a simulation based on real data using the Culture \& Community in a Time of Crisis dataset is presented in Section~\ref{sec: app}. Concluding  remarks are given in Section~\ref{sec: conc}.

\section{Problem Setup}\label{sec: problem setup}

Let $\mathcal{F}_N = \{(\bx_i,y_i):i=1,\ldots,N\}$ be a finite population of size $N$, where $\bx_i$ is a $p\geq 1$ dimensional  covariate vector associated with unit $i$, and $y_i$ is the response of interest. In this paper, we adopt a model-assisted framework \citep{sarndal1992} and assume that the finite population $\mathcal{F}_N$ is a random sample from a super-population model $F(\bx,y)$. The finite population is treated as fixed once it is generated, and we are interested in estimating the population mean $\theta_N=N^{-1}\sum_{i=1}^Ny_i$.

Due to the limited budget and time, it is practically impossible to observe $\{y_i:i=1,\ldots,N\}$. Instead, complex sampling is always performed to obtain a probability sample from the finite population $\mathcal{F}_N$,  and the size of the probability sample is usually much smaller compared to $N$. Thus, we assume that a probability sample $\{(\bx_i,y_i):i\in A\}$ of size $n_A$ is available, with $\pi_
{Ai}$ being the inclusion probability associated with the unit $i$ and $n_A = o_p(N)$.  Other than the probability sample $A$, we also assume the availability of an independent non-probability sample $\{(\bx_i,y_i):i\in B\}$ of size $n_B$ as well as the auxiliary information $\{\bx_i:i=1,\ldots,N\}$. Since the response of interest is observable in the probability sample,  $\hat{\theta}_A = N^{-1}\sum_{i\in A}\pi_{Ai}^{-1}y_i$ is a design-unbiased estimator of $\theta_N$ \citep{horvitz1952}. 
Denote $\pi_{Bi} = \Pr(\delta_{Bi}=1\mid \bx_i,y_i)$ to be the selection probability of the $i$th unit for the non-probability sample $B$. If $\{\pi_{Bi}:i\in B\}$ were fully observed, then we can obtain another design-unbiased estimator $\hat{\theta}_B = N^{-1}\sum_{i\in B}\pi_{Bi}^{-1}y_i$. Due to the independence between the probability sample $A$ and the non-probability sample $B$, we can construct a more efficient estimator based on $\hat{\theta}_A$ and $\hat{\theta}_B$, and the auxiliary information can be further incorporated. However,  the selection probabilities  $\{\pi_{Bi}:i\in B\}$  are not available in practice, and they need to be estimated to correct the selection bias associated with the non-probability sample $B$.





In this paper, we develop methods to combine the information from two samples as well as $\{\bx_i:i=1,\ldots,N\}$. 
To address the challenges and utilize all available information in statistically defensible ways, it is important to develop better statistical tools for data integration. 
 Many existing methods \citep{valliant2011estimating,chen2020doubly,wang2021adjusted} are mainly based on the MAR assumption  of \citet{rubin1976}, which is a strong assumption and difficult to validate in practice.  How to avoid or relax this assumption and develop valid statistical tools for data integration and data fusion  is a critically important problem in survey sampling.

\section{Maximum empirical likelihood estimator}\label{sec: MELE}
Consider the following MNAR response model for the non-probability sample: 
\begin{eqnarray}
    \pi_{Bi} = \pi(\bx_i,y_i;\bphi_0), \label{eq: true response model section 3}
\end{eqnarray}
where $\bphi_0$ is the  model parameter. Since a non-probability sample usually corresponds to a big data source, it is reasonable to assume $n_B=O_p(N)$. Rather than assuming the availability of the exact parametric form (\ref{eq: true response model section 3}) in this paper, we borrow the idea of multiple robustness \citep{han2013} and assume that the correct response model (\ref{eq: true response model section 3}) is contained in $K$ candidates $\{\pi_{k}(\bx,y;\bphi_k):k=1,\ldots,K\}$, where $\pi_{k}(\bx,y;\bphi_k)$ is the $k$th candidate response model with model parameter $\bphi_k$. The estimator $\hat{\bphi}_k$  solves 
         \begin{equation}
          \sum_{i=1}^N \left\{ \frac{ \delta_{Bi}}{\pi_k( \bx_i, y_i; \bphi_k) } -1 \right\} \bx_i = \mathbf{0}\label{eq: est phi}
         \end{equation}
for $k=1,\ldots,K$.
Other estimating equations can be used as long as certain conditions are satisfied; see \citet{Beaumont2005}, \citet{kim2007nonresponse}, \citet{chen2020doubly} and Section~7.3 of \citet{kim2021statistical} for details.

Based on estimated candidate models $\{\pi_{k}(\bx,y;\hat{\bphi}_k):k=1,\ldots,K\}$, we propose an approach based on empirical likelihood \citep{owen1991empirical} to combine the information from the probability sample $A$, the nonprobability sample $B$, and the auxiliary information $\{\bx_i:i=1,\ldots,N\}$ in the finite population. 
Specifically, the proposed maximum empirical likelihood (MEL) estimator $\hat{\theta}_{MEL}$ is obtained by maximizing 
 \begin{equation}
             \sum_{i=1}^N \delta_{Bi}  \log p_{i} \notag
        \end{equation}
        subject to $\sum_{i=1}^{N} \delta_{Bi}{p_{i}}  = 1$ and
         \begin{eqnarray} 
            &\sum_{i=1}^{N} \delta_{Bi}{p_{i}}  \pi_{k}(\bx_{i},y_{i};\hat{\bphi}_{k})  =  N^{-1} \sum_{i=1}^N \tilde{\pi}_k (\bx_i; \hat{\bbeta},\hat{\bphi}_k)\quad(k=1,\ldots,K),\label{eq: response model calibration} \\ 
            &\sum_{i=1}^{N} \delta_{Bi}{p_{i}} (y_i-\theta)=0,\label{eq: other constraint}
            \end{eqnarray}
        where  $\tilde{\pi}_k (\bx; \bbeta,\bphi) = E\{\pi_{k}(\bX,Y;\bphi)\mid \bX=\bx; \bbeta\}$. The right-hand side of (\ref{eq: response model calibration}) is a consistent estimator of the population mean $N^{-1}\sum_{i=1}^N\pi_k(\bx_i,y_i)$. Including the propensity score function in the calibration constraint in the empirical likelihood framework has been discussed by \cite{liu2023} and \cite{kim2023}.  
        The multiple calibration constraints in (\ref{eq: response model calibration}) for $\{p_i,i\in B\}$ intend to correct the selection bias of the non-probability sample and thus are referred to as multiple bias-calibration.  To obtain an estimator of  $\tilde{\pi}_k (\bx;\bbeta, \bphi_k)$, we need to assume a correctly specified conditional density function $f(y\mid \bx;\bbeta)$ with parameter $\bbeta$, and its estimator $\hat{\bbeta}$  is obtained by maximizing the pseudo log-likelihood function 
         \begin{equation}
         \sum_{i\in A}\pi_{Ai}^{-1}\log f(y_i\mid \bx_i;\bbeta).\label{eq: beta est}
         \end{equation}
 Since we do not restrict ourselves to the MAR assumption and do not need to pinpoint which candidate model is the correct one among the $K$ candidate models, we need an additional assumption  that $f(y\mid\bx;\bbeta)$ is correct. This condition is weaker than the correct model assumption for (\ref{eq: true response model section 3}) for the following reasons. First, the model assumption $f(y\mid\bx;\bbeta)$ is testable using the probability sample $A$. If a correct parametric model is difficult to specify,  we can consider a non-parametric density model and use fractional imputation \citep{kim2014fractional} to obtain $\tilde{\pi}_k (\bx; \bbeta,\bphi)$. Second, it is well known that the MNAR assumption (\ref{eq: true response model}) is very hard to validate even coupled with the probability sample $A$ under general setups.

Our proposal differs from existing work in the following aspects.  Existing methods mainly assumed MAR response models for data integration \citep{elliott2017inference,chen2020doubly,wang2021adjusted,wang2022functional}, but in this article we consider an MNAR response model (\ref{eq: true response model section 3}) for data fusion.   Although such an MNAR response model is discussed in the literature of missing data analysis \citep{kim2021statistical}, we do not assume a correctly specified parametric form for it. Instead, we consider the case where the correct response model is contained in several candidate models \citep{han2013}. \citet{han2013} did not provide a consistent variance estimator, but we propose an asymptotically unbiased variance estimator for our estimator that allows for multiply robust inference; see Section~\ref{sec: var est} for details.

    In addition to the constraint on the selection probabilities (\ref{eq: response model calibration}), an MEL generalized regression (MEL\_GREG) estimator can be obtained using the following additional constraint, 
          \begin{eqnarray} 
            &\sum_{i=1}^{N} \delta_{Bi}{p_{i}}  \bx_i =  N^{-1} \sum_{i=1}^N \bx_i.\label{eq: equality constraint}
          \end{eqnarray}
    Such marginal information is commonly used to improve the estimation efficiency in survey sampling; see, e.g., \citet{deville1992calibration}.

    The following estimation procedure is used to obtain the MEL estimator.
\begin{enumerate}
    \renewcommand{\theenumi}{[Step \arabic{enumi}]}
    \item Estimate $\bphi_k$  by solving $\sum_{i=1}^N \left\{ \delta_{Bi}\pi_k^{-1}( \bx_i, y_i; \bphi_k)  -1 \right\} \bx_i = \mathbf{0}$ for $k=1,\ldots,K$ in (\ref{eq: est phi}).
    \item Posit a conditional density model $f(y\mid \bx;\bbeta)$ and estimate $\bbeta$ by maximizing $\sum_{i\in A}\pi_i^{-1}\log f(y_i\mid \bx_i;\bbeta)$ in (\ref{eq: beta est}).
    
    \item Maximize $\sum_{i\in B}\log p_i$
    subject to $\sum_{i\in B}p_i=1$ and (\ref{eq: response model calibration})--(\ref{eq: other constraint}).
\end{enumerate}
We can use the algorithm in \citet{wu2005algorithms} to solve Step 3. To obtain the MEL\_GREG estimator, we only need an additional equality constraint (\ref{eq: equality constraint}) in Step 3. Other estimating functions can be used for $\theta$ if we are interested in parameters other than the population mean.

\section{Variance estimation} \label{sec: var est}
Since we have assumed $n_A=o_p(N)$ and $n_B = O_p(N)$, and since the sampling indicators $\{\delta_{Bi}:i=1,\ldots, N\}$ are independently generated for the non-probability sample $B$, we conclude that the variability of estimating $\bphi_k$ is asymptotically negligible compared to the uncertainty of estimating $\bbeta$ under regularity conditions. Therefore, we use $\bphi_k$ to denote its estimator $\hat{\bphi}_k$ in this section for the variance estimation of $\hat{\theta}$ as well as $\hat{\bbeta}$. 

The proposed  MEL estimator is asymptotically equivalent to the one by maximizing
 \begin{equation}
             \sum_{i=1}^N \delta_{Bi}  \log p_{i} \label{appeq: max p}
\end{equation}
        subject to $\sum_{i=1}^{N} \delta_{Bi}{p_{i}}  = 1$ and
         \begin{eqnarray} 
            &\displaystyle\sum_{i=1}^{N} \delta_{Bi}{p_{i}}  \pi_{k}(\bx_{i},y_{i};{\bphi}_{k})  =  N^{-1} \sum_{i=1}^N \tilde{\pi}_k (\bx_i; \hat{\bbeta},{\bphi}_k)\quad(k=1,\ldots,K),\label{appeq: response model calibration} \\ 
            &\displaystyle\sum_{i=1}^{N} \delta_{Bi}{p_{i}} (y_i-\theta)=0,\label{appeq: est equation for theta}\\
            & \displaystyle\sum_{i=1}^N\delta_{Ai}\pi_{Ai}^{-1}\frac{1}{f(y_i\mid \bx_i;\hat{\bbeta})}\frac{\partial f(y_i\mid \bx_i;\hat{\bbeta})}{\partial \bbeta}=\bzero,\label{appeq: est for beta}
            \end{eqnarray}
where (\ref{appeq: est for beta}) corresponds to the pseudo-maximum likelihood estimator in (\ref{eq: beta est}). The difference between (\ref{eq: response model calibration}) and (\ref{appeq: response model calibration}) is that we use the true parameter $\phi_k$ for the latter.

Let $\bpi_i = (\pi_{1}(\bx_{i},y_{i};{\bphi}_{1}),\ldots,\pi_{K}(\bx_{i},y_{i};{\bphi}_{K}))^{\T}$ and $$\bar{\bpi}(\bbeta) = N^{-1}\sum_{i=1}^N(\tilde{\pi}_1 (\bx_i; {\bbeta},{\bphi}_1),\ldots,\tilde{\pi}_K (\bx_i; {\bbeta},{\bphi}_K))^{\T}.$$Then, by the Lagrange multiplier, the optimization problem (\ref{appeq: max p})--(\ref{appeq: est for beta}) is equivalent to solving the following equation system
\begin{equation}
    \displaystyle\bbg(\bbeta,\blambda,\theta) = \begin{pmatrix}
    \displaystyle\sum_{i\in A}\pi_{Ai}^{-1}\frac{1}{f(y_i\mid \bx_i;\bbeta)}\frac{\partial f(y_i\mid \bx_i;\bbeta)}{\partial \bbeta}\\
    \displaystyle\sum_{i\in B}\frac{1}{1+\blambda^{\T}(\bpi_i-\bar{\bpi}(\bbeta))}(\bpi_i-\bar{\bpi}(\bbeta))\\
    \displaystyle\sum_{i\in B}\frac{1}{1+\blambda^{\T}(\bpi_i-\bar{\bpi}(\bbeta))}(y_i-\theta)
    \end{pmatrix}=\bzero. \label{eq: estimating equation}
\end{equation}

Under regularity conditions \citep{owen1990empirical}, we can show that the solution to (\ref{eq: estimating equation}), denoted as $(\hat{\bbeta}, \hat{\blambda},\hat{\theta})$, satisfies 
\begin{equation}
    (\hat{\bbeta}, \hat{\blambda},\hat{\theta})-(\bbeta,\bzero, \theta_N) = o_p(1).\label{eq: consistency par}
\end{equation} 
Therefore, linearization can be used to obtain a variance estimator for the solution, which is 
$$\hat{V}(\hat{\bbeta}, \hat{\blambda},\hat{\theta}) = \hat{\bI}^{-1}\hat{\bV}_g(\hat{\bI}^{-1})^{\T},$$
where $\hat{\bV}_g$ is the design variance estimator of $\bbg(\bbeta,\blambda,\theta)$ with $(\bbeta,\blambda,\theta)$ replaced by its estimator, and $\hat{\bI}$ evaluated by $ \partial \bbg(\bbeta,\blambda,\theta)/\partial (\bbeta,\blambda,\theta)^{\T}$ with parameters replaced by their estimators.

As we have discussed in the preceding section, we can provide an asymptotically unbiased variance estimator for the proposed estimator, but \citet{han2013} cannot unless a true model is available. This merit is achieved based on the assumption $n_A=o_p(n_B)$. Since a parametric density model is assumed for the response of interest $y$ given the covariate $\bx$, based on the consistency result in (\ref{eq: consistency par}), we can also derive an asymptotic distribution for the parameter of interest using linearization for statistical inference.

\section{Simulation} \label{sec: simu section}
In this section, we compare the performance of the proposed estimator with its competitors numerically. A finite population $\{(\bx_i,y_i):i=1,\ldots,N\}$ is generated as follows, where $\bx_i=(x_{i1},x_{i2})^{\T}$. For $i=1,\ldots,N$, $x_{ij}\sim N(0,1)$ for $j=1,2$, and $y_i = x_{i1} + x_{i2} + \epsilon_i$ with $\epsilon_i\sim N(0,4)$, where $N(\mu,\sigma^2)$ is a normal distribution with mean $\mu$ and variance $\sigma^2$. The parameter of interest is the population mean $\theta_N = N^{-1}\sum_{i=1}^Ny_i$. In this section, we consider $N\in\{5\,000, 10\,000\}$.

A simple random sampling without replacement is performed to generate a probability sample $\{(\bx_i,y_i):i\in A\}$ of size $n_A$, and the associated inclusion probability is $\pi_{A,i}=n_AN^{-1}$ for $i\in A$. The inclusion probability is similar to the selection probability, but is used for the probability sample. We consider $n_A\in\{100,400\}$.  To generate a non-probability sample $\{(\bx_i,y_i):i\in B\}$, we use
$
\delta_i\sim\mbox{Bernoulli}(\pi_{Bi}(\bx_i,y_i))
$ to generate the corresponding response indicators, where $\mbox{Bernoulli}(p)$ is a Bernoulli distribution with success probability $p\in(0,1)$. We consider the following three scenarios for the response model (\ref{eq: true response model}):
\begin{enumerate}
    \item[S1. ] $\mbox{logit}(\pi_{Bi}) = -0.5 + 0.5x_{i1} + 0.5x_{i2}$, where $\mbox{logit}(p) =\log p-\log(1-p)$ for $p\in(0,1)$.
    \item[S2. ] $\mbox{logit}(\pi_{Bi}) = -0.5 + 0.5x_{i1} + 0.5x_{i2}+0.2y_i$.
    \item[S3. ] $\mbox{logit}(\pi_{Bi}) = -0.5 + 0.5x_{i1} + (0.5x_{i2}+0.2y_i)I(y_i<0) + 0.3y_iI(y_i\geq0)$.
\end{enumerate}
For the three scenarios, the average response rates are about 0.4. Scenarios S1 and S2 correspond to the MAR assumption and the MNAR assumption, respectively. Additionally, scenario S3 is also an MNAR response model, but it cannot be correctly estimated using a traditional logistic regression model $\mbox{logit}(\pi_{Bi}) = \phi_0+\phi_1x_{i1}+\phi_2x_{i2}+\phi_3y_{i}$. 

We consider the following methods to estimate $\theta_N$.
\begin{enumerate}
    \item The HT estimator $\hat{\theta}_{HT} = n_A^{-1}\sum_{i\in A}y_i$  only by the probability sample $A$.
    \item The Generalized regression (GREG) estimator $\hat{\theta}_{GREG}=N^{-1}\sum_{i=1}^N\bx_i^{\T}\hat{\bgamma}$  incorporating information from the probability sample $A$ as well as auxiliary information $\{\bx_i:i=1,\ldots,N\}$, where $\hat{\bgamma} = (\sum_{i\in A}\pi_{Ai}^{-1}\bx_i\bx_i^{\T})^{-1}\sum_{i\in A}\pi_{Ai}^{-1}y_i\bx_i$.
	\item The RDW estimator \citep{valliant2011estimating}. Assume a logistic regression model for the selection probability associated with the non-probability sample $\pi_{Bi} = \pi_{Bi}(\bphi)$:
	\begin{equation}
		\log\left\{\frac{ \pi_{Bi}(\bphi)}{1- \pi_{Bi}(\bphi)}\right\} = \tilde{\bx}_i^{\top}\bphi, \label{eq: logistic model}
	\end{equation}
	where $\bphi$ is the model parameter to be estimated and $\tilde{\bx}_i^{\top} = (1,\bx_i^{\top})$ is a vector of covariates augmented with an intercept term. Parameter $\bphi$ is estimated by maximizing 
	\begin{equation}
	l_{RDW}(\bphi) = \sum_{i\in B}w_i^*\log  \pi_{Bi}(\bphi) + \sum_{i\in A}\log\{1 -  \pi_{Bi}(\bphi)\},\label{eq: l rdw}
	\end{equation}
	where $w_i^* = 1$ for $i\in B$, $w_i^* = \pi_{A,i}^{-1}(\hat{N}-n_B)/\hat{N}$ for $i\in A$,  and $\hat{N} = \sum_{i\in S_p}\pi_{A,i}^{-1}$ is the estimated population size. Denote $\hat{\bphi}$ as the one maximizing (\ref{eq: l rdw}), and let $w_i^{(RDW)} = \{\pi_{Bi}(\hat{\bphi})\}^{-1}$. Then, the RDW estimator is $$\hat{\theta}_{RDW} = \left(\sum_{i\in B}w_i^{(RDW)}\right)^{-1}\sum_{i\in B}w_i^{(RDW)}y_i.$$
	\item The Chen-Li-Wu (CLW) estimator \citep{chen2020doubly}. Consider (\ref{eq: logistic model}), but the model parameter $\bphi$ is estimated by maximizing 
	\begin{equation}
		l_{CLW}(\bphi) = \sum_{i\in B}	\log\left\{\frac{ \pi_{Bi}(\bphi)}{1- \pi_{Bi}(\bphi)}\right\} + \sum_{i\in A}\pi_{A,i}^{-1}\log\{1-\pi_{Bi}(\bphi)\}.
	\end{equation}
	Once the estimator $\hat{\bphi}$ is obtained,  a similar estimator to the RDW one can be used to estimate $\theta_N$.
	\item The adjusted logistic propensity weighting (ALP) estimator \citep{wang2021adjusted}. Denote $p_i = \pi_{Bi}(1-\pi_{Bi})^{-1}$, and consider a logistic regression model for $p_i = p_i(\bphi)$,
	\begin{equation}
			\log\left\{\frac{ p_i(\bphi)}{1- p_i(\bphi)}\right\} = \tilde{\bx}_i^{\top}\bphi. \label{eq: p logistic model}
	\end{equation}
	The parameter $\bphi$ is estimated by maximizing 
	\begin{equation}
		l_{ALP}(\bphi) = \sum_{i\in B}\log p_i(\bphi) + \sum_{i\in A}\pi_{A,i}^{-1}\log\{1-p_i(\bphi)\}.\notag
	\end{equation}
	Once an estimator $\hat{\bphi}$ is obtained, the ALP estimator is $$\hat{\theta}_{ALP} =\left(\sum_{i\in B}w_i^{(ALP)}\right)^{-1}\sum_{i\in B}w_i^{(ALP)}y_i,$$ where $w_i^{(ALP)} = \{1-p_i(\hat{\bphi})\}/p_i(\hat{\bphi})$.
	\item The full design weight (FDW) estimator \citep{wang2021adjusted}. Consider the ALP estimator, but use $p_i(\hat{\bphi})$ to approximate $\pi_{Bi}$. That is, the FDW estimator is $$\hat{\theta}_{FDW} =\left(\sum_{i\in B}w_i^{(FDW)}\right)^{-1}\sum_{i\in B}w_i^{(FDW)}y_i ,$$ where $w_i^{(FDW)} = 1/p_i(\hat{\bphi})$.
	\item The scaled ALP (ALP\_s) estimator \citep{wang2021adjusted}. Consider (\ref{eq: p logistic model}) but scale the sampling weights by a factor $\lambda = n_B/(\sum_{i\in A}\pi_{A,i}^{-1})$. That is, the parameter in (\ref{eq: p logistic model})  is estimated by maximizing 
	\begin{equation*}
				l_{ALP\_s}(\bphi) = \sum_{i\in B}\log p_i(\bphi) + \lambda\sum_{i\in A}\pi_{A,i}^{-1}\log\{1-p_i(\bphi)\}.
	\end{equation*}
	Let $\hat{\bphi}^{\top} =(\hat{\phi}_{0,\lambda},\hat{\bphi}_{1,\lambda}^{\top})$, and denote $w_{i}^{(ALP\_s)} = \exp(\bx_i^{\top}\hat{\bphi}_{1,\lambda})$. Then, the ALP\_s estimator is $$
	\hat{\theta}_{ALP\_s} = \left(\sum_{i\in B}w_i^{(ALP\_s)}\right)^{-1}\sum_{i\in B}w_i^{(ALP\_s)}y_i.
	$$
    \item The traditional EL (EL\_0) estimator, and it is obtained by maximizing 
    \begin{equation}
             \sum_{i=1}^N \delta_{Bi}  \log p_{i} \notag
        \end{equation}
        subject to $\sum_{i=1}^{N} \delta_{Bi}{p_{i}}  = 1$, $\sum_{i=1}^{N} \delta_{Bi}{p_{i}}\bx_i  = N^{-1}\sum_{i=1}^N\bx_i$, and $\sum_{i=1}^{N} \delta_{Bi}{p_{i}} (y_i-\theta)=0$.
    \item The EL\_1 estimator. First, estimate the density model parameter $\bbeta$ by maximizing (\ref{eq: beta est}).
    Then consider the following response model 
    $
        \mathrm{logit}\,\pi_1(\bx,y;{\bphi_1}) = {\phi_{10}} + {\phi_{11}} x_{1} + {\phi_{12}}x_{2},
    $
    where $\bphi_1 = (\phi_{10},\phi_{11},\phi_{12})'$, and it is estimated by solving (\ref{eq: est phi}).
    Then, maximize 
        $
             \sum_{i=1}^N \delta_i  \log {p_{i}} 
        $
        subject to $\sum_{i=1}^{N} \delta_{i}{p_{i}}  = 1$,
        $
            \sum_{i=1}^{N} \delta_{i}{p_{i}}  \pi_{1}(\bx_{i},y_{i};\hat{\bphi}_{1})  =  N^{-1} \sum_{i=1}^N \tilde{\pi}_1 (\bx_i; \hat{\bphi}_1)$, and $\sum_{i=1}^{N} \delta_{i}{p_{i}} (y_i-\theta)=0,
        $
        where $\tilde{\pi}_1 (\bx; \bphi) = E\{\pi_{1}(\bx_{i},y_{i};\bphi;\hat{\bbeta})\}$. 
    \item The EL\_2 estimator, which is similar to EL\_1 estimator, but we consider the following response model 
    $
        \mathrm{logit}\,\pi_2(\bx,y;{\bphi_2})  = {\phi_{20}} + {\phi_{21}} x_{1} + {\phi_{22}}y,
    $
    where $\bphi_2 = (\phi_{20},\phi_{21},\phi_{22})'$.
    \item The EL\_3 estimator, which is similar to EL\_1 estimator, but we consider the following response model 
    $
        \mathrm{logit}\,\pi_3(\bx,y;{\bphi_3})  = {\phi_{30}} + {\phi_{31}} x_{2} + {\phi_{32}}y,
    $
    where $\bphi_3 = (\phi_{30},\phi_{31},\phi_{32})'$.
    \item The proposed MEL estimator based on the estimated parameters $\hat{\bbeta}$, $\hat{\bphi}_1$, $\hat{\bphi}_2$ and $\hat{\bphi}_3$ associated with the above EL\_1, EL\_2 and EL\_3 estimators.
    \item The proposed MEL\_GREG estimator with an additional constraint $ \sum_{i=1}^{N} \delta_{i}{p_{i}}  \bx_i =  N^{-1} \sum_{i=1}^N \bx_i$.
\end{enumerate}
The HT estimator and the GREG estimator are commonly used in practice. Other existing estimators, including the RDW estimator, CLW estimator, ALP estimator, FDW estimator, ALP\_s estimator, are based on the MAR assumption for the response model. In this section, we consider three candidate models by collecting  those for EL\_1 to EL\_3 estimators. For Scenario S1, one of the candidate response models is correctly specified, but all three are wrong for the other two scenarios. Thus, Scenarios S2--S3 can be used to check the robustness of the proposed MEL and MEL\_GREG estimators.

\begin{figure}
    \centering
\includegraphics[width=\textwidth]{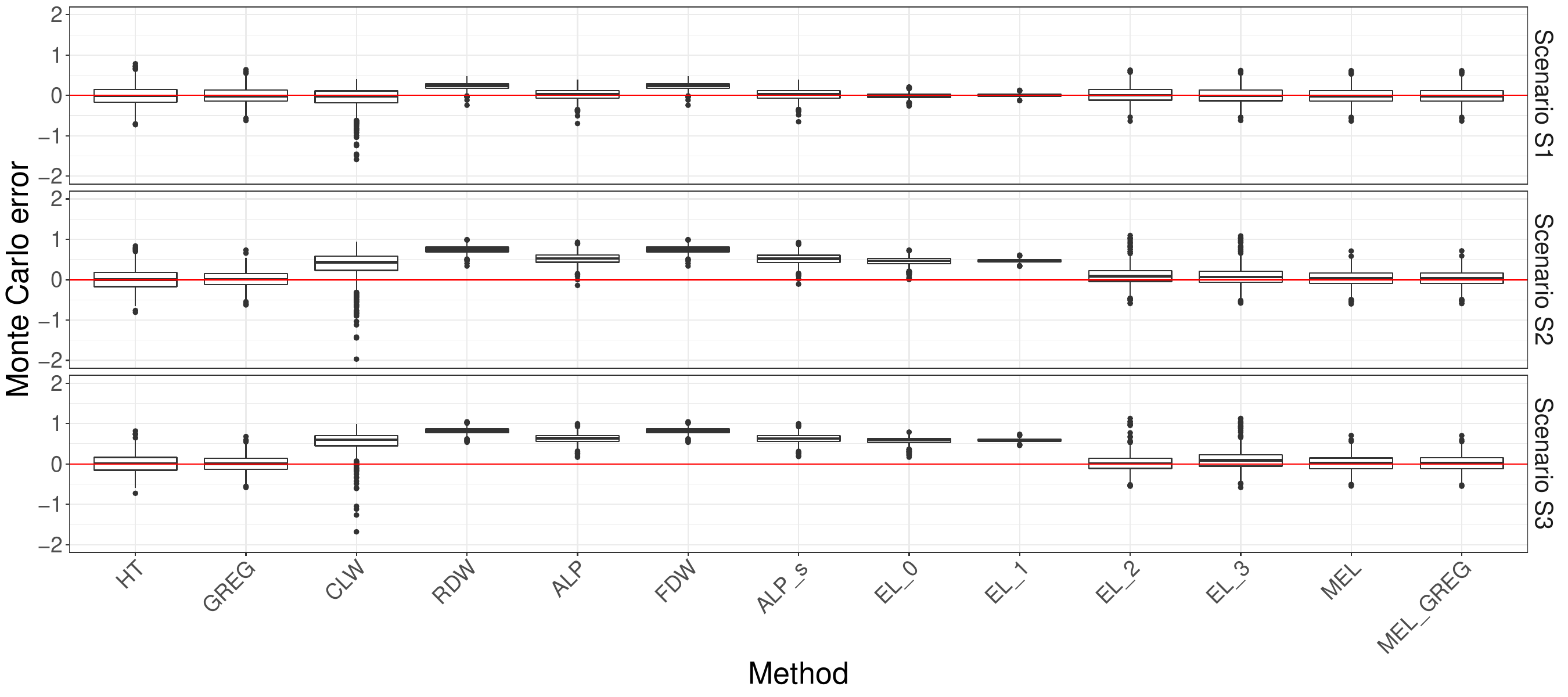}
    \caption{Comparison of different estimators under the setup $N=5\,000$ and $n_A=100$.}
    \label{fig:all estimator1}
\end{figure}

Based on 1\,000 Monte Carlo simulations, Figure~\ref{fig:all estimator1} shows the comparison of different estimators for the three different scenarios under the setup $N=5\,000$ and $n_A=100$. When the response model satisfies the MAR assumption, the RDW estimator and the FDW estimator are slightly biased and all remaining ones are consistent. Among the consistent estimators, the CLW estimator is not as stable as others, since its variability is greater. The EL\_1 estimator with correctly specified response model outperforms others in terms of variability. The EL\_0 estimator performs similarly as the EL\_1 estimator but with slightly more variability. However, when the MAR assumption fails for Scenarios S2--S3, most existing estimators are biased, except for the HT estimator, the GREG estimator, the EL estimators, and the two proposed MEL estimators. The same observation holds for different setups with different population and sample sizes. To simplify the discussion, we compare only the HT estimator, the GREG estimator, the three EL estimators, and the two proposed estimators in the following analysis.

Table~\ref{tab: MSE} shows the summary statistics for different estimators under different setups based on 1\,000 Monte Carlo simulations. It is obvious that the HT estimator is less efficient than the GREG estimator, since the latter incorporates auxiliary information. Other than the conclusions about EL\_1 estimator from Figure~\ref{fig:all estimator1}, EL\_2 estimator and EL\_3 estimator perform poorly under Scenarios S2--S3, regardless of the sizes for the finite population and the probability sample.   The proposed MEL and MEL\_GREG estimators is more efficient than the GREG estimator in general, and the MEL\_GREG estimator is slightly more efficient than the MEL estimator, since additional auxiliary information is incorporated. If the sample size $n_A$ remains unchanged, the bias, variance, and MSE are comparable for different population sizes, and such an observation is guaranteed by the theoretical assumption $n_A = o_p(N)$.
\begin{table}[!ht]
\centering
\caption{Summary statistics for different estimators under different scenarios based on 1\,000 Monte Carlo simulations. The unit for ``Bias'' is $10^{-2}$, and it is $10^{-4}$ for ``Var'' and ``MSE'', where ``Var'' stands for variance, and ``MSE'' for mean squared error.} \label{tab: MSE}
\begin{tabular}{cccrrrrrrrrrrr}
  \hline
\multirow{2}{*}{$N$} & \multirow{2}{*}{$n_A$} & \multirow{2}{*}{Method} & \multicolumn{3}{c}{Scenario 1}&&\multicolumn{3}{c}{Scenario 2}&&\multicolumn{3}{c}{Scenario 3}\\
&&&Bias & Var & MSE &  & Bias & Var & MSE  &  &Bias & Var & MSE  \\ 
  \hline
\multirow{15}{*}{$5000$} & \multirow{7}{*}{$100$} & HT & -1 & 611 &  612 &  &  0 & 658 &  658 &  &  1 & 581 &  581 \\ 
   &  & GREG &  0 & 397 &  397 &  &  1 & 424 &  424 &  &  0 & 426 &  426 \\ 
   &  &EL\_1 &  0 &  17 &   \textbf{17} &  & 47 &  20 & 2183 &  & 59 &  17 & 3488 \\ 
   &  & EL\_2 &  2 & 389 &  392 &  &  9 & 506 &  595 &  &  2 & 493 &  498 \\ 
   &  & EL\_3 &  0 & 391 &  391 &  &  8 & 539 &  597 &  & 10 & 543 &  641 \\ 
   &  & MEL & -1 & 383 &  383 &  &  3 & 377 &  384 &  &  1 & 386 &  388 \\ 
   &  & MEL\_GREG &  0 & 383 &  383 &  &  3 & 373 &  \textbf{381} &  &  2 & 382 &  \textbf{385} \\ 
   &  &  &  &  &  &  &  &  &  &  &  &  &  \\ 
   & \multirow{7}{*}{$400$} & HT &  0 & 138 &  138 &  &  0 & 138 &  138 &  &  0 & 137 &  137 \\ 
   &  & GREG &  0 &  86 &   86 &  &  0 &  92 &   92 &  &  0 &  94 &   94 \\ 
   &  &EL\_1 &  0 &  \textbf{18} &   18 &  & 46 &  21 & 2156 &  & 59 &  18 & 3503 \\ 
   &  & EL\_2 &  2 &  91 &   94 &  &  8 & 217 &  287 &  &  2 & 162 &  165 \\ 
   &  & EL\_3 &  0 &  87 &   87 &  &  7 & 244 &  295 &  &  9 & 191 &  272 \\ 
   &  & MEL &  0 &  88 &   88 &  &  2 &  85 &   88 &  &  1 &  89 &   91 \\ 
   &  & MEL\_GREG &  0 &  85 &   85 &  &  2 &  83 &  \textbf{ 87} &  &  2 &  87 &   \textbf{90} \\ 
   &  &  &  &  &  &  &  &  &  &  &  &  &  \\ 
   &  &  &  &  &  &  &  &  &  &  &  &  &  \\ 
  \multirow{15}{*}{$10000$} & \multirow{7}{*}{$100$} & HT &  0 & 602 &  602 &  &  1 & 513 &  513 &  & -1 & 566 &  567 \\ 
   &  & GREG &  1 & 396 &  396 &  &  1 & 361 &  362 &  &  0 & 387 &  387 \\ 
   &  & EL\_1 &  0 &   8 &    \textbf{8} &  & 47 &  11 & 2187 &  & 60 &   9 & 3584 \\ 
   &  & EL\_2 &  2 & 379 &  384 &  & 11 & 595 &  719 &  &  3 & 570 &  581 \\ 
   &  & EL\_3 &  2 & 379 &  384 &  &  8 & 398 &  469 &  &  9 & 417 &  496 \\ 
   &  & MEL &  1 & 377 &  378 &  &  4 & 318 &  \textbf{333} &  &  2 & 359 &  362 \\ 
   &  & MEL\_GREG &  1 & 375 &  376 &  &  4 & 315 &  \textbf{333} &  &  2 & 357 &  \textbf{360} \\ 
   &  &  &  &  &  &  &  &  &  &  &  &  &  \\ 
   & \multirow{7}{*}{$400$} & HT &  0 & 145 &  145 &  &  0 & 145 &  145 &  &  0 & 144 &  144 \\ 
   &  & GREG &  0 &  97 &   98 &  &  0 &  94 &   94 &  &  0 &  97 &   97 \\ 
   &  &EL\_1 &  0 &   8 &    \textbf{8} &  & 47 &  10 & 2184 &  & 60 &   9 & 3607 \\ 
   &  & EL\_2 &  2 &  97 &  100 &  & 11 & 354 &  471 &  &  4 & 327 &  346 \\ 
   &  & EL\_3 &  2 &  95 &   98 &  &  8 & 138 &  195 &  &  9 & 139 &  218 \\ 
   &  & MEL &  0 &  93 &   93 &  &  3 &  82 &   \textbf{91} &  &  2 &  88 &   \textbf{92} \\ 
   &  & MEL\_GREG &  1 &  93 &   93 &  &  3 &  93 &  105 &  &  2 &  88 &   \textbf{92} \\ 
   \hline
\end{tabular}
\end{table}

Table~\ref{tab: var est} shows the relative bias of the variance estimator under different setups based on 1\,000 Monte Carlo simulations, and it is calculated as 
$
(\bar{V} - V)/V,
$
where $V$  is the variance of a certain estimator,  $\bar{V} =1000^{-1}\sum_{m=1}^{1000}\hat{V}_m$, $\hat{V}_m$ is the variance estimator of this estimator for the $m$th simulation study. We use the Monte Carlo variance to approximate $V$.
When the population size is $N=5\,000$, the relative bias of the variance estimator is satisfactory under Scenario S1, since its absolute values are less than 0.05 regardless of the values of $n_A$. However, the relative biases are large for both Scenarios S2--S3. In addition, the variance estimator underestimates the variance for most setups, and one possible reason is that the finite population size is not so large that the variability due to the non-probability sample is not negligible. As the finite population size increases to $N=10\,000$,  the performance of the variance estimator improves in general.
\begin{table}[!ht]
\centering
\caption{Relative bias of the variance estimator under different setups based on 1\,000 Monte Carlo simulations.}\label{tab: var est}
\begin{tabular}{ccrrrrrrrr}
  \hline
\multirow{2}{*}{$N$} & \multirow{2}{*}{$n_A$}& \multicolumn{2}{c}{Scenario S1}&&\multicolumn{2}{c}{Scenario S2}&&\multicolumn{2}{c}{Scenario S3}\\
&&MEL & MEL\_GREG &  &MEL & MEL\_GREG &  & MEL & MEL\_GREG \\ 
  \hline
\multirow{2}{*}{$5000$} & $100$ & -0.019 & -0.028 &  & -0.113 & -0.105 &  & -0.092 & -0.090 \\ 
   & $400$ & -0.009 & 0.026 &  & -0.084 & -0.057 &  & -0.081 & -0.070 \\ 
   &  &  &  &  &  &  &  &  &  \\ 
  \multirow{2}{*}{$10000$} & $100$ & -0.002 & -0.001 &  & 0.064 & 0.072 &  & -0.014 & -0.009 \\ 
   & $400$ & -0.024 & -0.020 &  & -0.020 & -0.017 &  & -0.035 & -0.036 \\ 
   \hline
\end{tabular}
\end{table}

Table~\ref{tab: cr} shows the coverage rates of a Wald two-sided 95\% confidence interval under different configurations based on 1\,000 Monte Carlo simulations. When the finite population size is $N=5\,000$, the coverage rates are smaller than its nominal truth of 0.95 for Scenarios S2--S3, since the true variance is underestimated. However, as the finite population size increases to $N=10\,000$, the coverage rates are close to 0.95, showing the satisfactory performance of the confidence intervals.
\begin{table}[!ht]
\centering
\caption{Coverage rates under different setups based on 1\,000 Monte Carlo simulations.}\label{tab: cr}
\begin{tabular}{ccrrrrrrrr}
  \hline
\multirow{2}{*}{$N$} & \multirow{2}{*}{$n_A$}& \multicolumn{2}{c}{Scenario S1}&&\multicolumn{2}{c}{Scenario S2}&&\multicolumn{2}{c}{Scenario S3}\\
&&MEL & MEL\_GREG &  &MEL & MEL\_GREG &  & MEL & MEL\_GREG \\ 
  \hline
\multirow{2}{*}{$5000$} & $100$ & 0.950 & 0.952 &  & 0.931 & 0.936 &  & 0.932 & 0.931 \\ 
   & $400$ & 0.947 & 0.946 &  & 0.944 & 0.950 &  & 0.941 & 0.941 \\ 
   &  &  &  &  &  &  &  &  &  \\ 
  \multirow{2}{*}{$10000$} & $100$ & 0.956 & 0.959 &  & 0.949 & 0.949 &  & 0.949 & 0.950 \\ 
   & $400$ & 0.950 & 0.949 &  & 0.938 & 0.939 &  & 0.951 & 0.950 \\ 
   \hline
\end{tabular}
\end{table}

\section{Culture \& Community in a Time of Crisis}\label{sec: app}
In this section, we compare various estimators using data from a real 2020 study,  Culture and Community in a Time of Crisis (CCTC), which surveyed more than 120,000 Americans online and collected information on hundreds of behavioral and attitudinal variables related to arts and culture before and during the pandemic \citep{benoit2021exploring}. Due to its large sample size, the CCTC dataset can be used to compare the performance of different data integration methods. 


In this section, we treat the 113,549 respondents as a finite population, and there are 293 variables with high response rates. Among those, we are interested in estimating the average response to the question ``Before Covid-19, how important or unimportant were arts \& culture organizations to you''. There are five scales to this question and 1--5 represent ``Very Unimportant'', ``Unimportant'', ``Neither'', ``Important'' and ``Very Important'', respectively. There are no non-responses for this question, and its population mean is 4.438. In addition to the population average, we are also interested in estimating the average response to this question among different age groups, and the domain means are 4.420, 4.406, 4.434, and 4.460 for age groups 1--4, respectively, where age groups 1--4 represent those aged 18--34, 35--49, 50--64 and older than 64. In addition to the variable of interest, we consider two regional variables as covariates.

We consider a simulation based on real data, and simple stratified random sampling without replacement is performed to obtain a probability sample of size 1\,000, where the sample size is proportional to the stratum size with a minimum number of 40; see \citet{benoit2021exploring} for more details. The following  is used to generate the response indicators for a non-probability sample:
\begin{equation}
    \mbox{logit}(\pi_{Bi}) = -2 +  0.2x_{i2}+0.3y_i, \label{eq: true response model}
\end{equation}
where $\bx_i=(x_{i1}, x_{i2})$ are the two regional covariates, and $y_i\in\{1,2,3,4,5\}$ is the response of interest associated with the $i$th person in the finite population. The average response rate is 0.375. Since the response of interest contains only five levels, we consider a multinomial distribution with linear effect for $f(y\mid\bx;\bbeta)$. We compare those estimators in Section~\ref{sec: simu section} based on 1\,000 Monte Carlo simulations. 

\subsection{Population average estimation}
\begin{figure}
    \centering
\includegraphics[width=\textwidth]{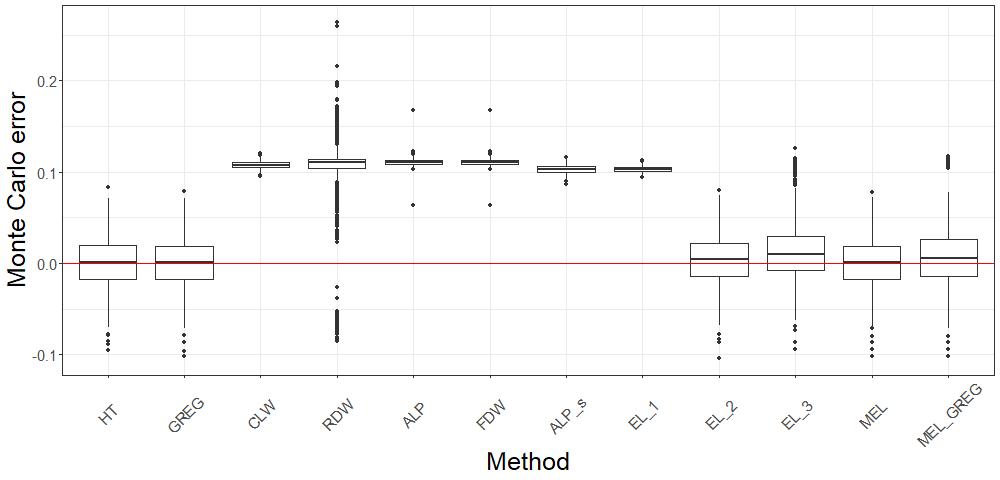}
    \caption{Comparison of different estimators based on 1\,000 Monte Carlo simulations using the CCTC dataset.}
    \label{fig:all estimator1 application}
\end{figure}

Figure~\ref{fig:all estimator1 application} shows the comparison of different estimators for the average attitude of the population toward art and cultural organizations. First, existing estimators, including CLW estimator, ALP estimator, FDW estimator and ALP\_s estimator, are biased since the response model is MNAR. The EL\_1 estimator is also biased due to the fact that it assumes an MAR model as other existing estimators. The EL\_2 and EL\_3 estimators are approximately unbiased, but, interestingly, the EL\_2 estimator performs better. Please note that the response model for the  EL\_3 estimator is correctly specified, but the response model for the EL\_2 estimator is wrong. One possible reason for this consequence may be that $x_{i2}$ only has two different levels, that is, $x_{i2}\in\{0,1\}$, but there are five different levels for the response of interest, that is,   $y_i\in\{1,2,3,4,5\}$. Furthermore, the coefficient for $x_{i2}$ is lower than that for the response of interest. Therefore, compared to $y_i$, $x_{i2}$ plays a far less important role in (\ref{eq: true response model}). We checked the two singular values of the design matrix and the first is much larger than the second one, indicating that there exists a certain collinearity between the two covariates. Furthermore, $x_{i1}\in\{1,2,3,4\}$, so it is more informative than $x_{i2}$. 
The proposed MEL and MEL\_GREG estimators are almost as efficient as the HT and GREG estimators when estimating the population average.

\subsection{Domain average estimation}
Other than population average, we are also interested in estimating the domain averages. In this subsection, we are interested in estimating the average attitude toward art and cultural organizations among different age groups. 

\begin{figure}
    \centering
\includegraphics[width=\textwidth]{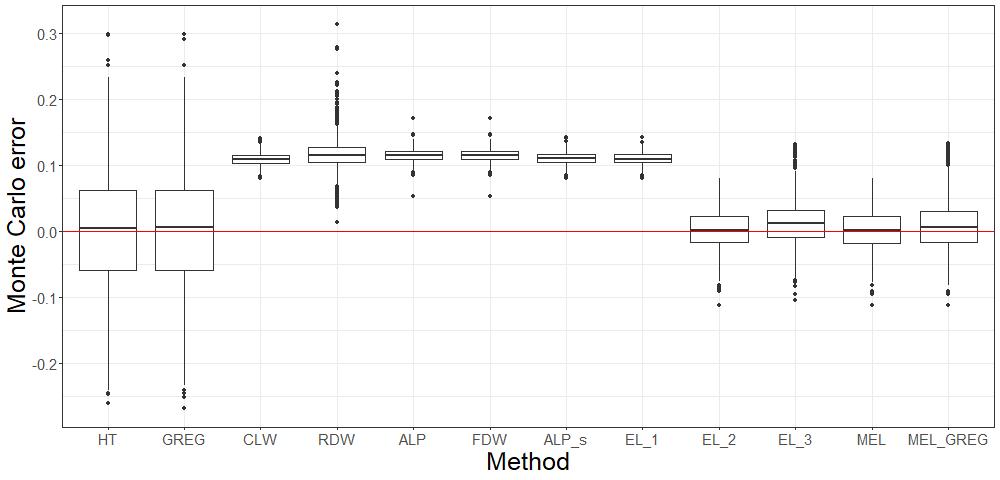}
    \caption{Comparison of different estimators based on 1\,000 Monte Carlo simulations using the CCTC dataset for age group 18--34.}
    \label{fig:all estimator1 application age1}
\end{figure}

Figure~\ref{fig:all estimator1 application age1} shows the Monte Carlo errors of different estimators associated with age group 18--34. Existing estimators as well as the EL\_1 estimator are biased, since they do not adjust the selection bias. Although the proposed MEL and MEL\_GREG estimators perform similarly to the HT and GREG estimators when estimating the population average, they are more efficient. We can draw the same conclusion for the other three age groups. This observation is not theoretically justified, but it would be an interesting research topic in the future. 
\section{Conclusion} \label{sec: conc}
We propose an MEL estimator and an MEL\_GREG estimator to integrate information from a probability sample, a non-probability sample and auxiliary information from the finite population, and  the associated variance estimators are also discussed. The proposed estimators allow for MNAR response models, but a correct conditional outcome density model given the auxiliary information should be specified. The simulation study shows the good performance of the proposed estimators, and an additional plasmode simulation study based on real data illustrates that the proposed estimators outperform the existing ones, especially in terms of domain estimation.   

It is possible to relax the requirement of a correct conditional outcome density model parametrically by resorting to flexible semi-parametric or nonparametric models. However, this will change the asymptotic properties and variance estimators of the MEL and MEL\_GREG estimators. We will pursue this topic as a future research.   

\bibliographystyle{chicago}
 \bibliography{ref1}

\begin{thebibliography}{}

\bibitem[\protect\citeauthoryear{Beaumont}{Beaumont}{2005}]{Beaumont2005}
Beaumont, J.-F. (2005).
\newblock Calibrated imputation in surveys under a quasi-model-assisted
  approach.
\newblock {\em Journal of the Royal Statistical Society. Series B (Statistical
  Methodology)\/}~{\em 67\/}(3), 445--458.
\newblock DOI: https://doi.org/10.1111/j.1467-9868.2005.00511.x.

\bibitem[\protect\citeauthoryear{Beaumont}{Beaumont}{2020}]{beaumont2020probability}
Beaumont, J.-F. (2020).
\newblock Are probability surveys bound to disappear for the production of
  official statistics.
\newblock {\em Survey Methodology\/}~{\em 46\/}(1), 1--28.

\bibitem[\protect\citeauthoryear{Benoit-Bryan and Mulrow}{Benoit-Bryan and
  Mulrow}{2021}]{benoit2021exploring}
Benoit-Bryan, J. and E.~Mulrow (2021).
\newblock Exploring nonprobability methods with simulations from a common data
  source: culture and community in a time of crisis.
\newblock In {\em To appear: Joint Statistical Meetings 2021 Proceedings}.

\bibitem[\protect\citeauthoryear{Bethlehem}{Bethlehem}{2016}]{bethlehem2016solving}
Bethlehem, J. (2016).
\newblock Solving the nonresponse problem with sample matching?
\newblock {\em Social Science Computer Review\/}~{\em 34}, 59--77.
\newblock DOI: https://doi.org/10.1177/0894439315573926.

\bibitem[\protect\citeauthoryear{Chen, Li, Rao, and Wu}{Chen
  et~al.}{2022}]{chen2022pseudo}
Chen, Y., P.~Li, J.~N.~K. Rao, and C.~Wu (2022).
\newblock Pseudo empirical likelihood inference for nonprobability survey
  samples.
\newblock {\em Canadian Journal of Statistics\/}~{\em 50\/}(4), 1166--1185.
\newblock DOI: https://doi.org/10.1002/cjs.11708.

\bibitem[\protect\citeauthoryear{Chen, Li, and Wu}{Chen
  et~al.}{2020}]{chen2020doubly}
Chen, Y., P.~Li, and C.~Wu (2020).
\newblock Doubly robust inference with nonprobability survey samples.
\newblock {\em Journal of the American Statistical Association\/}~{\em
  115\/}(532), 2011--2021.
\newblock DOI: https://doi.org/10.1080/01621459.2019.1677241.

\bibitem[\protect\citeauthoryear{Citro}{Citro}{2014}]{citro2014multiple}
Citro, C.~F. (2014).
\newblock From multiple modes for surveys to multiple data sources for
  estimates.
\newblock {\em Survey Methodology\/}~{\em 40\/}(2), 137--162.

\bibitem[\protect\citeauthoryear{Couper}{Couper}{2000}]{couper2000web}
Couper, M.~P. (2000).
\newblock Review: Web surveys: A review of issues and approaches.
\newblock {\em The Public Opinion Quarterly\/}~{\em 64\/}(4), 464--494.
\newblock DOI: https://doi.org/10.1086/318641.

\bibitem[\protect\citeauthoryear{Couper}{Couper}{2013}]{couper2013sky}
Couper, M.~P. (2013).
\newblock Is the sky falling? new technology, changing media, and the future of
  surveys.
\newblock In {\em Survey Research Methods}, Volume~7, pp.\  145--156.
\newblock DOI: https://doi.org/10.18148/srm/2013.v7i3.5751.

\bibitem[\protect\citeauthoryear{Deville and S{\"a}rndal}{Deville and
  S{\"a}rndal}{1992}]{deville1992calibration}
Deville, J.-C. and C.-E. S{\"a}rndal (1992).
\newblock Calibration estimators in survey sampling.
\newblock {\em Journal of the American statistical Association\/}~{\em
  87\/}(418), 376--382.
\newblock DOI: https://doi.org/10.2307/2290268.

\bibitem[\protect\citeauthoryear{Elliott and Valliant}{Elliott and
  Valliant}{2017}]{elliott2017inference}
Elliott, M. and R.~Valliant (2017).
\newblock Inference for nonprobability samples.
\newblock {\em Statistical Science\/}~{\em 32\/}(2), 249--264.
\newblock DOI: https://doi.org/10.1214/16-STS598.

\bibitem[\protect\citeauthoryear{H{\'a}jek}{H{\'a}jek}{1971}]{hajek1971comment}
H{\'a}jek, J. (1971).
\newblock Comment on “an essay on the logical foundations of survey sampling,
  part one”.
\newblock {\em The Foundations of Survey Sampling\/}~{\em 236}.

\bibitem[\protect\citeauthoryear{Han and Wang}{Han and Wang}{2013}]{han2013}
Han, P. and L.~Wang (2013).
\newblock Estimation with missing data: Beyond double robustness.
\newblock {\em Biometrika\/}~{\em 100}, 417--430.
\newblock DOI: https://doi.org/10.1093/biomet/ass087.

\bibitem[\protect\citeauthoryear{Holt}{Holt}{2007}]{holt2007official}
Holt, D.~T. (2007).
\newblock The official statistics olympic challenge: Wider, deeper, quicker,
  better, cheaper.
\newblock {\em The American Statistician\/}~{\em 61\/}(1), 1--8.
\newblock DOI: https://doi.org/10.1198/000313007X168173.

\bibitem[\protect\citeauthoryear{Horvitz and Thompson}{Horvitz and
  Thompson}{1952}]{horvitz1952}
Horvitz, D.~G. and D.~J. Thompson (1952).
\newblock A generalization of sampling without replacement from a finite
  universe.
\newblock {\em Journal of the American Statistical Association\/}~{\em 42},
  663--685.

\bibitem[\protect\citeauthoryear{Kalton}{Kalton}{2019}]{kalton2019developments}
Kalton, G. (2019).
\newblock Developments in survey research over the past 60 years: A personal
  perspective.
\newblock {\em International Statistical Review\/}~{\em 87}, S10--S30.
\newblock DOI: https://doi.org/10.1111/insr.12287.

\bibitem[\protect\citeauthoryear{Kim and Kim}{Kim and
  Kim}{2007}]{kim2007nonresponse}
Kim, J.~K. and J.~J. Kim (2007).
\newblock Nonresponse weighting adjustment using estimated response
  probability.
\newblock {\em Canadian Journal of Statistics\/}~{\em 35\/}(4), 501--514.
\newblock DOI: https://doi.org/10.1002/cjs.5550350403.

\bibitem[\protect\citeauthoryear{Kim and Morikawa}{Kim and
  Morikawa}{2023}]{kim2023}
Kim, J.~K. and K.~Morikawa (2023).
\newblock An empirical likelihood approach to reduce selection bias in
  voluntary samples.
\newblock Avaialbe at arXiv.2211.02998.

\bibitem[\protect\citeauthoryear{Kim, Park, Chen, and Wu}{Kim
  et~al.}{2021}]{kim2021combining}
Kim, J.~K., S.~Park, Y.~Chen, and C.~Wu (2021).
\newblock Combining non-probability and probability survey samples through mass
  imputation.
\newblock {\em Journal of the Royal Statistical Society Series A: Statistics in
  Society\/}~{\em 184\/}(3), 941--963.
\newblock DOI: https://doi.org/10.1111/rssa.12696.

\bibitem[\protect\citeauthoryear{Kim and Shao}{Kim and
  Shao}{2021}]{kim2021statistical}
Kim, J.~K. and J.~Shao (2021).
\newblock {\em Statistical Methods for Handling Incomplete Data\/} (2nd ed.).
\newblock CRC press.

\bibitem[\protect\citeauthoryear{Kim and Tam}{Kim and Tam}{2021}]{kimtam2021}
Kim, J.-K. and S.-M. Tam (2021).
\newblock Data integration by combining big data and survey sample data for
  finite population inference.
\newblock {\em International Statistical Review\/}~{\em 89\/}(2), 382--401.
\newblock DOI: https://doi.org/10.1111/insr.12434.

\bibitem[\protect\citeauthoryear{Kim and Yang}{Kim and
  Yang}{2014}]{kim2014fractional}
Kim, J.~K. and S.~Yang (2014).
\newblock Fractional hot deck imputation for robust inference under item
  nonresponse in survey sampling.
\newblock {\em Survey Methodology\/}~{\em 40}, 211--230.

\bibitem[\protect\citeauthoryear{Liu and Fan}{Liu and Fan}{2023}]{liu2023}
Liu, Y. and Y.~Fan (2023).
\newblock Biased-sample empirical likelihood weighting for missing data
  problems: an alternative to inverse probability weighting.
\newblock {\em Journal of the Royal Statistical Society: Series B\/}~{\em 85},
  67--83.

\bibitem[\protect\citeauthoryear{Meng}{Meng}{2018}]{meng2018statistical}
Meng, X.-L. (2018).
\newblock Statistical paradises and paradoxes in big data (i) law of large
  populations, big data paradox, and the 2016 {US} presidential election.
\newblock {\em Annals of Applied Statistics\/}~{\em 12\/}(2), 685--726.
\newblock DOI: https://doi.org/10.1214/18-AOAS1161SF.

\bibitem[\protect\citeauthoryear{Miller}{Miller}{2017}]{miller2017there}
Miller, P.~V. (2017).
\newblock Is there a future for surveys?
\newblock {\em Public Opinion Quarterly\/}~{\em 81\/}(S1), 205--212.
\newblock DOI: https://doi.org/10.1093/poq/nfx008.

\bibitem[\protect\citeauthoryear{Owen}{Owen}{1990}]{owen1990empirical}
Owen, A. (1990).
\newblock Empirical likelihood ratio confidence regions.
\newblock {\em Annals of Statistics\/}~{\em 18\/}(1), 90--120.
\newblock DOI: https://doi.org/10.1214/aos/1176347494.

\bibitem[\protect\citeauthoryear{Owen}{Owen}{1991}]{owen1991empirical}
Owen, A. (1991).
\newblock Empirical likelihood for linear models.
\newblock {\em The Annals of Statistics\/}, 1725--1747.
\newblock DOI: https://doi.org/10.1214/aos/1176348368.

\bibitem[\protect\citeauthoryear{Pfefffermann}{Pfefffermann}{2015}]{pfeffermann15}
Pfefffermann, D. (2015).
\newblock Methodological issues and challenges in the production of official
  statistics: 24th {A}nnual {M}orris {H}ansen {L}ecture.
\newblock {\em Journal of the Survey Statistics and Methodology\/}~{\em 3},
  425--483.
\newblock DOI: https://doi.org/10.1093/jssam/smv035.

\bibitem[\protect\citeauthoryear{Rao}{Rao}{2020}]{rao2020}
Rao, J.~N.~K. (2020).
\newblock On making valid inferences by integrating data from surveys and other
  sources.
\newblock {\em Sankhya B\/}.
\newblock Accepted (DOI: https://doi.org/10.1007/s13571-020-00227-w).

\bibitem[\protect\citeauthoryear{Rivers}{Rivers}{2007}]{rivers2007sampling}
Rivers, D. (2007).
\newblock Sampling for web surveys.
\newblock In {\em Joint Statistical Meetings}, Volume~4. American Statistical
  Association Alexandria, VA.

\bibitem[\protect\citeauthoryear{Rubin}{Rubin}{1976}]{rubin1976}
Rubin, D.~B. (1976).
\newblock Inference and missing data.
\newblock {\em Biometrika\/}~{\em 63\/}(3), 581--592.
\newblock DOI: https://doi.org/10.1093/biomet/63.3.581.

\bibitem[\protect\citeauthoryear{S\"{a}rndal}{S\"{a}rndal}{1992}]{sarndal1992}
S\"{a}rndal, C.~E. (1992).
\newblock Methods for estimating the precision of survey estimates when
  imputation is used.
\newblock {\em Survey Methodology\/}~{\em 18}, 241--252.

\bibitem[\protect\citeauthoryear{Tam and Clarke}{Tam and Clarke}{2015}]{tam15}
Tam, S.-M. and F.~Clarke (2015).
\newblock Big data, official statistics and some initiatives by the
  {A}ustralian {B}ureau of {S}tatistics.
\newblock {\em International Statistical Review\/}~{\em 83}, 436--448.
\newblock DOI: https://doi.org/10.1111/insr.12105.

\bibitem[\protect\citeauthoryear{Valliant and Dever}{Valliant and
  Dever}{2011}]{valliant2011estimating}
Valliant, R. and J.~A. Dever (2011).
\newblock Estimating propensity adjustments for volunteer web surveys.
\newblock {\em Sociological Methods \& Research\/}~{\em 40\/}(1), 105--137.
\newblock DOI: https://doi.org/10.1177/0049124110392533.

\bibitem[\protect\citeauthoryear{Wang, Valliant, and Li}{Wang
  et~al.}{2021}]{wang2021adjusted}
Wang, L., R.~Valliant, and Y.~Li (2021).
\newblock Adjusted logistic propensity weighting methods for population
  inference using nonprobability volunteer-based epidemiologic cohorts.
\newblock {\em Statistics in Medicine\/}~{\em 40\/}(24), 5237--5250.
\newblock DOI: https://doi.org/10.1002/sim.9122.

\bibitem[\protect\citeauthoryear{Wang, Mao, and Kim}{Wang
  et~al.}{2022}]{wang2022functional}
Wang, Z., X.~Mao, and J.~K. Kim (2022).
\newblock Functional calibration under non-probability survey sampling.
\newblock Available at arXiv:2204.09193.

\bibitem[\protect\citeauthoryear{Williams and Brick}{Williams and
  Brick}{2018}]{williams2018trends}
Williams, D. and J.~M. Brick (2018).
\newblock Trends in us face-to-face household survey nonresponse and level of
  effort.
\newblock {\em Journal of Survey Statistics and Methodology\/}~{\em 6},
  186--211.
\newblock DOI: https://doi.org/10.1093/jssam/smx019.

\bibitem[\protect\citeauthoryear{Wu}{Wu}{2005}]{wu2005algorithms}
Wu, C. (2005).
\newblock Algorithms and r codes for the pseudo empirical likelihood method in
  survey sampling.
\newblock {\em Survey Methodology\/}~{\em 31\/}(2), 239.

\bibitem[\protect\citeauthoryear{Wu}{Wu}{2022}]{wu2022statistical}
Wu, C. (2022).
\newblock Statistical inference with non-probability survey samples.
\newblock {\em Survey Methodology\/}~{\em 48\/}(2), 283--311.

\bibitem[\protect\citeauthoryear{Yang and Kim}{Yang and Kim}{2020}]{yang2020}
Yang, S. and J.~Kim (2020).
\newblock Statistical data integration in survey sampling: A review.
\newblock {\em Japanese Journal of Statistics and Data Science\/}~{\em 3},
  625--650.

\end{thebibliography}
\end{document}